\documentclass[runningheads]{llncs}
\usepackage[T1]{fontenc}
\usepackage{lmodern}

\usepackage[numbers, sort, comma, square]{natbib}
\usepackage{graphicx}
\usepackage{multirow}%
\usepackage{amsmath,amssymb,amsfonts}%
\usepackage{csquotes}
\usepackage{mathrsfs}%
\usepackage[title]{appendix}%
\usepackage{textcomp}%
\usepackage{manyfoot}%
\usepackage{booktabs}%
\usepackage{algorithm}%
\usepackage{algorithmicx}%
\usepackage{algpseudocode}%
\usepackage{listings}%
\usepackage{graphicx} %
\usepackage{subcaption} %
\usepackage{etoolbox}
\usepackage{xspace}
\usepackage[english]{babel}
\usepackage{enumitem}
\usepackage{hyperref}
\usepackage{mdframed}
\usepackage{multirow}
\usepackage[dvipsnames]{xcolor}
\usepackage{tcolorbox}
\usepackage{caption}
\usepackage{newfloat}
\usepackage{subscript}
\usepackage{tablefootnote}

\newcommand{\llmprompt}[4]
{
\bigskip
    \noindent\begin{minipage}{\textwidth}
        \begin{tcolorbox}[
            colback=Apricot!5!white,
            colframe=green!30!black,
            fonttitle=\bfseries, 
            arc=1mm,
            boxrule=0.5mm,
            width=\textwidth,
            after=\captionof{figure}{#2}\label{#3}, 
            title={#1},
            ]
            #4
        \end{tcolorbox}
    \end{minipage}
    \vspace{.15cm}
}

\addto\extrasenglish{%
}

\newcommand{\llama}{\textit{Llama-3.1-Instruct-8B}\xspace}
\newcommand{\mistral}{\textit{Mistral-Instruct-7B}\xspace}
\newcommand{\gpt}{\textit{GPT-4o-mini}\xspace}
\newcommand{\claude}{\textit{Claude-3-5-sonnet}\xspace}
\newcommand{\sd}[1]{\textsubscript{$\pm$#1}}

\makeatletter
\newcommand{\printfnsymbol}[1]{%
  \textsuperscript{\@fnsymbol{#1}}%
}
\makeatother
\begin{document}

\title{Enhancing Tourism Recommender Systems for Sustainable City Trips Using Retrieval-Augmented Generation}
\titlerunning{Sustainable TRS for City Trips using RAG}

\author{Ashmi Banerjee\inst{1}\thanks{These authors contributed equally.} \and
Adithi Satish\inst{1}\printfnsymbol{1} \and
Wolfgang Wörndl\inst{1, 2}}

\institute{Technical University of Munich, Munich, Germany \\
\email{\{adithi.satish, ashmi.banerjee\}@tum.de} \and woerndl@in.tum.de}
\authorrunning{A. Satish et al.}

\authorrunning{Banerjee et al.}

\maketitle              %

\begin{abstract}
    Tourism Recommender Systems (TRS) have traditionally focused on providing personalized travel suggestions, often prioritizing user preferences without considering broader sustainability goals. 
    Integrating sustainability into TRS has become essential with the increasing need to balance environmental impact, local community interests, and visitor satisfaction. 
    This paper proposes a novel approach to enhancing TRS for sustainable city trips using Large Language Models (LLMs) and a modified Retrieval-Augmented Generation (RAG) pipeline. 
    We enhance the traditional RAG system by incorporating a sustainability metric based on a city’s popularity and seasonal demand during the prompt augmentation phase. 
    This modification, called Sustainability Augmented Reranking (SAR), ensures the system's recommendations align with sustainability goals.
    Evaluations using popular open-source LLMs, such as \llama and \mistral, demonstrate that the SAR-enhanced approach consistently matches or outperforms the baseline (without SAR) across most metrics, highlighting the benefits of incorporating sustainability into TRS.

\end{abstract}
\section{Introduction}

Tourism Recommender Systems (TRS) have been widely used to assist travelers by providing personalized suggestions for accommodations, activities, destinations, and more~\cite{ISINKAYE2015261}. 
Previously, these systems have largely been centered around the user's perspective, with their needs being the only criteria for their evaluation.
However, in recent times, they have evolved into platforms that must balance the interests of multiple stakeholders, creating a multistakeholder environment~\cite{abdollahpouri2020multistakeholder}. 
Tourism has far-reaching effects beyond its direct stakeholders, impacting the environment, local businesses, and residents. Thus, a TRS should incorporate sustainable options and practices to address these diverse effects and promote responsible tourism. 
This is particularly important in this domain, as it faces unique challenges such as seasonality, travel regulations, and limited resources like airline tickets and hotel availability~\cite{balakrishnan2021multistakeholder}.

The United Nations World Tourism Organization defines sustainable tourism as "\textit{tourism that takes full account of its current and future economic, social and environmental impacts, addressing the needs of visitors, the industry, the environment, and host communities}"~\cite{gossling2017tourism}.
The significance of sustainable tourism development has become increasingly evident, especially with the growing threat of climate change~\cite{anne_hardy_sustainable_2002, butler1999sustainable, liu2003sustainable, weaver2007sustainable}, emphasizing the need to integrate sustainability into TRS.
Although much work has been exploring the multistakeholder nature of TRS~\cite{rahmaniPOI,shen2021sar,weydemann2019defining,wu2021tfrom}, work focusing on generating sustainable recommendations is limited~\cite{banerjee_review_2023}.

Recent studies have increasingly explored the potential of Large Language Models (LLMs) to provide personalized user recommendations~\cite{liu2023chatgpt, dai2023uncovering}.
However, given the dynamic nature of tourism data, a TRS using LLM must adapt to frequent updates and changes. 
Fine-tuning LLMs to address these changes is unfeasible and resource-intensive due to the substantial computational costs and time needed for each update~\cite{rajbhandari2020zero}.
Moreover, LLMs are prone to hallucinations, where they provide answers that contradict real-world knowledge or are irrelevant to user prompts~\cite{zhang2023siren, ji2023survey}.
One effective way to address these challenges, is to implement a Retrieval-Augmented Generation (RAG) pipeline, which has proven successful in dynamic content scenarios~\cite{di2023retrieval, ding_retrieve_2024}. 
This approach enhances LLMs by integrating additional information from external databases~\cite{lewis2020retrieval}. 

This paper proposes a novel way to generate sustainable recommendations by leveraging LLMs using RAG.
We aim to recommend sustainable European cities based on natural language user queries on vacation planning.
To achieve this, we create a knowledge base of tourism information for 160 European cities, covering attractions, hotels, and restaurants.
We refine the traditional RAG system by incorporating a sustainability metric based on a city’s popularity and seasonal demand during the prompt augmentation phase. 
This enhancement, termed Sustainability Augmented Reranking (SAR), ensures that recommendations prioritize sustainability.
We evaluate our system with and without the SAR enhancement using two popular open-source LLMs, \llama and \mistral. 
Results show that SAR consistently matches or outperforms the baseline across most metrics, underscoring the efficacy of integrating sustainability into TRS.
Our approach helps mitigate hallucinations and allows for a multi-stakeholder perspective in recommendations, balancing user preferences with sustainability principles.

This paper is organized as follows:~\autoref{section: related} reviews the related work, ~\autoref{section: pipeline} describes our approach, ~\autoref{section: results} presents the results of our evaluation, comparing our method with and without the SAR enhancement using \llama and \mistral, and ~\autoref{section: conclusion} concludes the paper by summarizing the key findings and suggesting directions for future research.

\section{Related Work} \label{section: related}

In the tourism domain, research on recommender systems is primarily centred around the user's perspective, with their historical preferences used in tour recommendation and their subsequent evaluation. User attributes like location history, geo-tagged photographs, and check-in behavior have previously been used to identify Points of Interest (POIs) that align with user interests~\citep{takeuchi2006cityvoyager, lim2019tour, cheng2011personalized, lim2018personalized, majid2013context, lu2012personalized}. This section surveys the existing work across Retrieval-Augmented LLMs for Recommender Systems and Sustainable TRS.

\subsection{Retrieval-Augmented LLMs for Recommender Systems}

A new tangent of research in recommender systems has emerged over the past decade, which involves utilizing the generation and reasoning capability of LLMs to provide recommendations~\cite{abbasi2021tourism, dai_uncovering_2023, lin2024recommendersystemsbenefitlarge, bao2023tallrec, ji2024genrec, zheng2024adapting}. In tourism, fine-tuning remains the primary approach to enable LLMs to generate personalized recommendations that cater to the end-user's preferences~\citep{wei2024tourllm, li2024research}. 

To address the challenge of hallucinations in LLMs, recent studies focus on augmenting LLMs with the retrieved user history and find that this combination of retrieval and text generation shows comparable performance with multiple baselines~\citep{di2023retrieval,gao2023chat}. This approach of using the RAG architecture to enrich LLM prompts with user-item information enhances the LLM's reasoning and conversational abilities, improving the quality and veracity of its recommendations by providing suitable explanations from the context~\citep{wu2024coral, lu2021revcorereviewaugmentedconversationalrecommendation, wang2024empowering}.

\subsection{Sustainable Tourism Recommender Systems}
Findings by~\citet{banerjee_review_2023} reveal the recent emergence of a promising dimension of research that considers Society or the environment as one of the stakeholders in the recommendation process.
For example, ~\citet{merinov2023sustainability} incorporates sustainability into a multistakeholder recommender system by modeling an objective to rearrange tourist flows to prevent overcrowding. 
Other sustainability objectives considered in prior work include economic sustainability~\cite{patro_towards_2020}, food security~\cite{pachot2021multiobjective}, air pollution~\cite{herzog2019integrating}, eco-friendly accommodations~\cite{hoffmann2022measuring}, and carbon emissions of different modes of transportation~\cite{banerjee_modeling_2024}. 

In contrast to this state-of-the-art research in sustainable TRS, which predominantly involves formulating different sustainability objectives, the novelty of our research lies in taking advantage of the inherent abilities of LLMs to process textual information and generate explanations for each recommended destination.
This allows us to simultaneously consider the user's preferences and ensure that sustainability objectives are met.

\section{Approach} \label{section: pipeline}

\begin{figure}
\includegraphics[width=\textwidth]{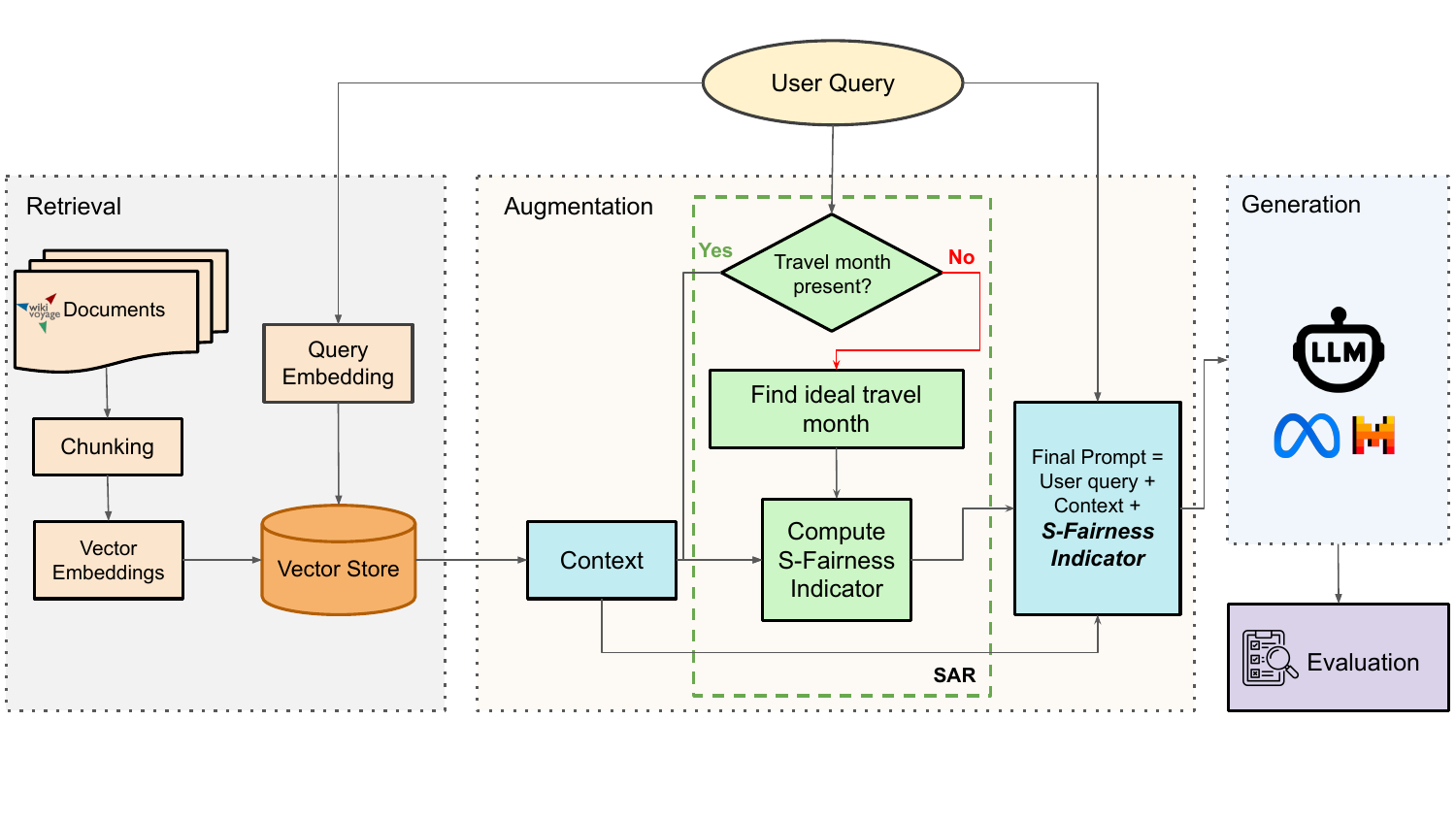}
\caption{Pipeline of our proposed modified RAG system. The \textcolor{ForestGreen}{green} dotted box shows the modified augmentation phase with the addition of the sustainability metric.\label{fig: pipeline}}
\end{figure}

We assume users visit our system seeking recommendations for European cities to travel to for vacation.
To effectively meet their needs, it is essential to implement a TRS that can adapt to the dynamic nature of tourism data, accommodating frequent updates and changes~\cite{balakrishnan2021multistakeholder,abdollahpouri2020multistakeholder}.
Continuously fine-tuning an LLM for these changes is not feasible, as it is inefficient and expensive to re-train the model every time the information is updated~\cite{rajbhandari2020zero}.
Therefore, we opt to use a modified Retrieval-Augmented Generation (RAG) pipeline as it demonstrates promising results in scenarios with frequently changing content~\cite{di2023retrieval}.

Typically, a naive RAG system comprises of three main phases --- information Retrieval, prompt Augmentation, and Generation~\cite{lewis2020retrieval}.
We modify the conventional naive RAG system by incorporating a sustainability metric (S-Fairness indicator) based on the popularity and the monthly seasonal demand of the city to re-rank the retrieved context during the prompt augmentation phase~\cite{banerjee_modeling_2024}.
This enhancement, termed Sustainability Augmented Reranking (SAR), ensures that the generated results prioritize sustainability considerations.
While \autoref{fig: pipeline} visually represents our system's workflow, the subsequent sections offer a detailed explanation of this process.

\subsection{Data Preparation}

For our knowledge base, we use data from Wikivoyage\footnote{https://www.wikivoyage.org/}, an online travel guide offering detailed information on travel destinations. Each city on Wikivoyage has a dedicated article with extensive details on transportation, weather, and tourist attractions, organized into various headings and subheadings. We specifically utilize two datasets from Wikivoyage: the \textit{Articles} and \textit{Listings} datasets.

\begin{figure}[ht!]
    \vspace{-3mm}
    \centering
    \begin{subfigure}[b]{0.4\textwidth} %
        \centering
        \includegraphics[scale=0.22]{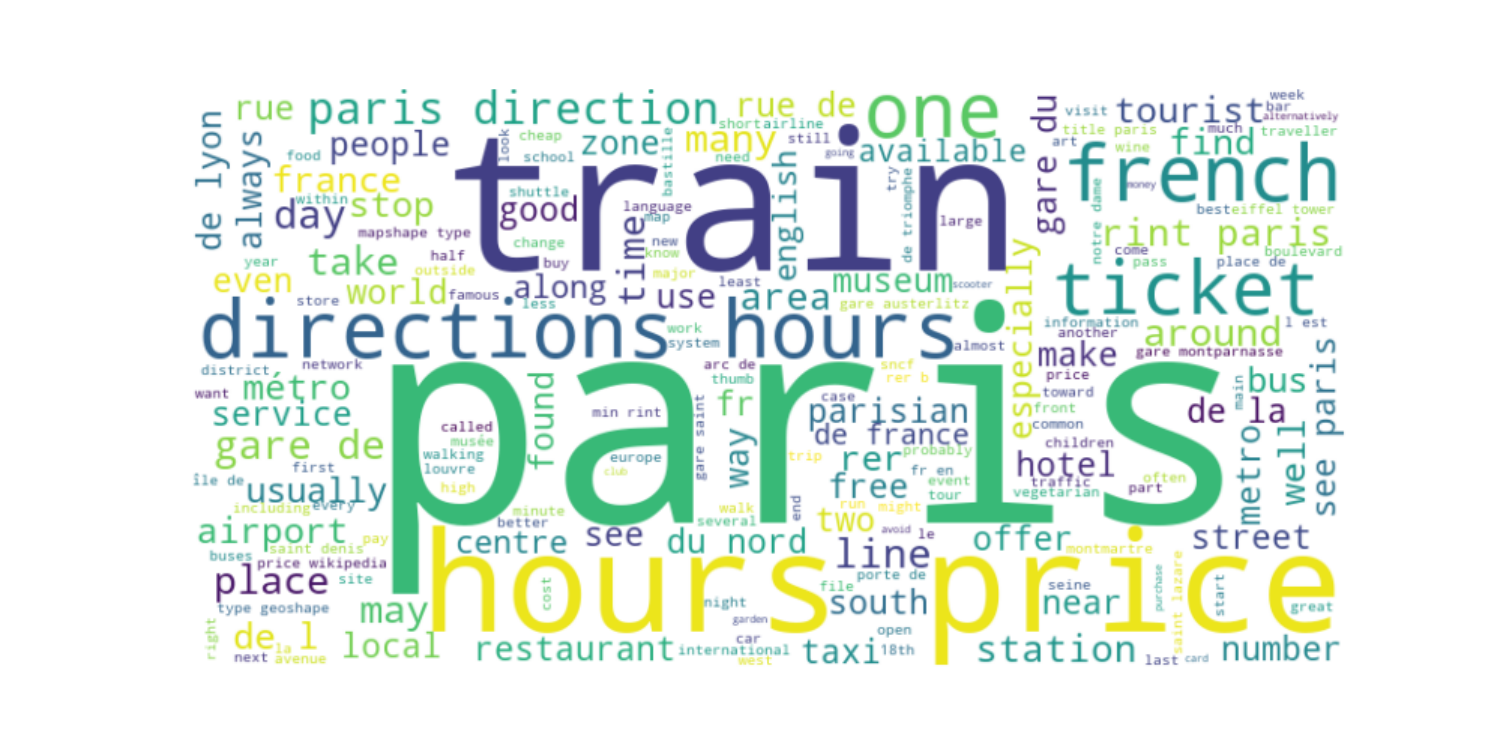}
        \caption{Wordcloud of the Wikivoyage Article page for Paris}
        \label{fig:paris_wordcloud} %
    \end{subfigure}%
    \hfill %
    \begin{subfigure}[b]{0.55\textwidth}
        \centering
        \includegraphics[scale=0.27]{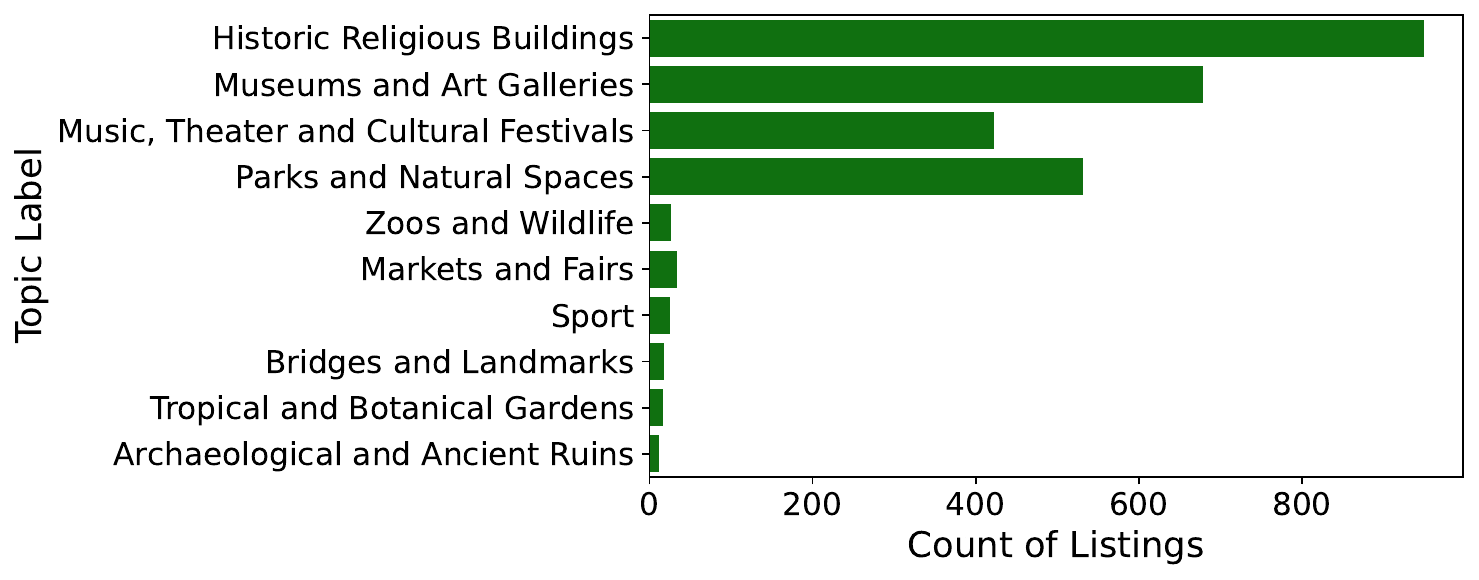}
        \caption{Counts of the Topic Clusters of POIs in the Wikivoyage Listings dataset}
        \label{fig:topic_counts} %
    \end{subfigure}
    \caption{Exploratory Data Analysis for Wikivoyage Articles and Listings}
    \label{fig:eda_wikivoyage} %
    \vspace{-5mm}
\end{figure}

The \textit{Articles}\footnote{https://dumps.wikimedia.org/enwikivoyage} dataset consists of XML files containing the full text of articles for various cities on Wikivoyage. It provides a broad textual overview of each city, highlighting key tourist-related information and city characteristics.
For instance,~\autoref{fig:paris_wordcloud} presents a word cloud of the most common terms in the Paris article from our dataset. Frequent terms such as "train," "price," "ticket," "hours," "directions," "louvre," "eiffel tower", "notre dame", and "museum" highlight the essential tourist details and reflect the city's unique features.

The \textit{Listings}\footnote{https://github.com/baturin/wikivoyage-listings} dataset, on the other hand, offers detailed descriptions of POIs, accommodations, and dining options. For our study, we focus on European cities, selecting a curated list of 160 cities across 41 countries from this dataset.
As illustrated in \autoref{fig:topic_counts}, the top 10 clusters generated using BERTopic~\cite{grootendorst2022bertopic} encompass a diverse range of POI categories, including culture, religion, natural landscapes, and wildlife from the \textit{Listings} dataset.

Together, these datasets provide a comprehensive textual landscape of travel destinations by combining general overviews with more detailed information about the cities.

To calculate the popularity and seasonality indices for the sustainability metric, described in~\autoref{section: prompt_augmentation}, we use data from Tripadvisor API~\footnote{https://tripadvisor-content-api.readme.io/reference/overview} to determine the aggregated number of POIs in each of the 160 cities. 
Monthly visitor footfall data is based on publicly available information from the \url{https://www.whereandwhen.net/} website. 
These counts are then normalized on a scale from 0 to 1 using Min-Max normalization technique~\cite{patro2015normalization} to establish each city's relative popularity and seasonality.

In our data, London ranks highest in popularity, followed by Paris, Rome, and Barcelona. 
The overall trend in seasonality indicates that most European cities experience high footfall during the summer months, although some cities exhibit specific spikes at particular times. 
For instance, Munich shows a significant increase in visitor counts in September, coinciding with Oktoberfest. 
Conversely, Brussels, a city with many business travelers~\cite{santos2018tourism}, has low seasonality indices during the summer months (May-August), a period typically associated with vacation in Europe.

\subsection{Information Retrieval} \label{section: info_retrieval}

The Information Retrieval phase in a RAG system involves extracting data from a large repository and storing it as embeddings in a vector database to provide context for generating accurate responses. 
In this paper, we utilize LanceDB\footnote{https://lancedb.github.io/lancedb/}, an open-source vector database, to store these embeddings and compute similarities during retrieval.
To efficiently manage document storage and ensure the retrieved context fits within the LLM's context window, we chunk Wikivoyage documents by subheadings, dividing them into smaller, logically coherent pieces to enhance retrieval and maintain contextual relevance~\cite{yepes2024financial}.

To compute embeddings, we utilize the \texttt{all-MiniLM-L6-V2} model\footnote{\url{https://huggingface.co/sentence-transformers/all-MiniLM-L6-v2}}, which leverages the pre-trained MiniLM model to map text into a 384-dimensional dense vector space~\cite{wang2020minilmdeepselfattentiondistillation}. 
Using this model, we generate embeddings for the chunked documents and transform the user query from natural language into embeddings. Cosine similarity~\cite{wikipedia_cosine_similarity} is then used to measure the similarity between the query embedding and the embeddings of the chunked documents. 
The top ten most similar cities with their respective chunks, which serve as the context, are subsequently returned.

\subsection{Sustainability Augmented Reranking (SAR)} \label{section: prompt_augmentation}

After retrieving the context, the next step involves augmenting the user query with this context. The novelty in our approach lies in incorporating sustainability considerations during this phase.
The sustainability of a destination is represented by its S-Fairness, or Societal Fairness, which reflects the equitable distribution of tourism's benefits and impacts on the non-participating stakeholders (Society) such as environment, residents and local businesses, involved in the recommendation process~\cite{ashmi_umap_dc_2023}. 

To compute this metric, we use a simplified approach based on a normalized weighted combination of popularity and seasonality indices for each destination and travel month, based on our prior work~\cite{banerjee_modeling_2024}.
We adapt the equation for calculating the sustainability metric $\psi(c_i^j)$, or S-Fairness indicator, for a destination city $c_i$ in month $j$ as the following:

\begin{align}\label{equation: s-fairness}
\psi(c_i^j) = 0.334 \cdot \rho(c_i) + 0.385 \cdot \sigma(c_i^j)
\end{align}
Where $\rho(c_i)$ indicates the popularity and $\sigma(c_i^j)$ seasonality indices, respectively. 
The weight coefficients are derived from the original work~\cite{banerjee_modeling_2024}, where we conducted a user survey to assess the importance of various factors influencing travel destination choices.
A lower $\psi(c_i^j)$ value indicates a more sustainable option for the given month. 
If the user does not specify a travel month, we calculate the S-Fairness indicator for the month with the lowest seasonality index for each destination retrieved in the context, and recommend that month as the ideal time to visit.

\llmprompt{System Prompt for Augmentation}{System prompt used for augmentation. The text in \textcolor{Sepia}{Sepia} represents the default prompt (without sustainability scoring information), and the text in \textcolor{ForestGreen}{green} represents the additional sustainability information to include SAR.}{fig: prompt}{
        \ttfamily\raggedright
        \textcolor{Sepia}{
            You are an AI recommendation system.\\ 
            Your task is to recommend European cities for travel based on the user's question. You should use the provided contexts to suggest the city best suited to the user's question. \\
            You recommend a list of the top three most sustainable cities to the user and the best month of travel. If the user has already provided the month of travel in the question, use the same month; otherwise, provide the ideal month. \\
            \textcolor{ForestGreen}{ A sustainable city is defined as a city with low overall popularity and low footfall for the intended month of travel. Each recommendation should also explain why it is being recommended on sustainability grounds. The context contains a sustainability score for each city, also known as the S-Fairness indicator, along with the ideal month of travel. A lower S-Fairness value indicates that the city is a better destination for the month provided. A city without a sustainability score should not be considered. You should only consider the S-Fairness indicator values while choosing the best city. However, your answer should not contain the numeric score itself. \\}
            Your answer must begin with "I recommend, " followed by the city name and why you recommended it. Your answers are correct, high-quality, and written by a domain expert. If the provided context does not contain the answer, state, "The provided context does not have the answer."
            }
}

Thus, the S-Fairness indicator aims to recommend more sustainable cities by considering their popularity and monthly visitor counts, thereby balancing tourist flow throughout the year and addressing issues of over- and under-tourism. 
This paper focuses exclusively on the impact of popularity and seasonality on sustainability, in contrast to the original model, which also accounted for emissions from travel to the destination.
This adjustment was made due to challenges in interpreting natural language queries to determine users' starting points and the limited availability of relevant data on public transport. 
However, the coefficients from the original model are retained as they were normalized in the initial equation.

We define a city's popularity by the number of POIs it has normalized between 0 and 1 across all cities, while seasonality refers to the normalized monthly footfall for the city.
Although the popularity index provides a general measure, it may not reflect unique characteristics or visitor trends in specific months. Therefore, with its monthly granularity, the seasonality index offers a more accurate depiction of crowd levels, ensuring that recommendations include less popular yet appealing and less crowded destinations~\cite{banerjee_modeling_2024}.

Once the sustainability data is added to the context, we instruct the LLM to balance user preferences (as outlined in the query) with sustainability concerns. 
We aim to determine whether including explicit sustainability information in the context and prompt leads to more sustainable recommendations than when the LLM provides recommendations without explicit sustainability scores.
We use two prompts to achieve this: one with explicit sustainability information (SAR) and one without (baseline).
We employ a "role-playing" prompting technique~\cite{jin2023lending}, where the system describes the role and responsibilities of the LLM, along with the user query and retrieved context. 
\autoref{fig: prompt} illustrates the prompts used for both scenarios.
For the SAR prompt, explicit sustainability information is included, as shown in \autoref{fig: prompt} in \textcolor{ForestGreen}{green}, whereas the baseline prompt omits this information. 
By comparing the outcomes from these two prompts, we aim to assess the impact of incorporating explicit sustainability information on the quality and sustainability of the recommendations provided by the LLM.

\subsection{Response Generation} \label{section: response_generation}

The final stage of a RAG pipeline involves response generation by the LLM, which uses the augmented prompt with the retrieved context to produce an output.
Here, the LLM performs two tasks --- selecting the three most suitable and sustainable cities based on the user query and generating an appropriate response using the provided context. 
It functions as a mediator in a multistakeholder scenario, balancing user preferences with sustainability concerns in its responses. Furthermore, we instruct the LLM to explain each city chosen in its recommendation. This capability of LLMs to generate human-like explanations improves the overall quality and transparency of the recommendations compared to more traditional methods. Finally, to evaluate this pipeline, we conduct a comparative analysis of the answer generation using two different open-source instruction-tuned LLMs, \llama~and \mistral, as described in~\autoref{section: results}, for both the SAR and baseline scenarios.

\section{Evaluation} \label{section: results}
In this section, we evaluate the effectiveness of our proposed approach by comparing it against a baseline method, i.e., without the SAR enhancement.
 We utilize two popular open-source instruction-tuned LLMs --- \llama~\cite{dubey2024llama} and \mistral~\cite{jiang2023mistral} to conduct this comparison, aiming to assess the impact of incorporating SAR on the quality of the generated recommendations. 
~\autoref{section: setup} describes our experimental setup, including the use of other LLMs to generate synthetic test cases for evaluation. In ~\autoref{section: metrics}, we discuss the different metrics used for evaluation and present the corresponding results.

\subsection{Setup: Synthetic Test Case Generation} \label{section: setup}

\llmprompt{Example Prompt and Model Response}{Example of a prompt generated by \gpt and responses generated by \llama. The response contains the top choice recommended by the model for the \textcolor{BlueViolet}{\textbf{baseline}} and \textcolor{ForestGreen}{\textbf{SAR}}. The top recommendation for SAR, Strasbourg, has the lowest S-Fairness indicator among the ten retrieved cities, indicating that the model effectively balances user preferences with sustainability concerns.}{fig: sample_prompt_response}{
\ttfamily 
\textbf{Prompt:}\\
\textcolor{Sepia}{\textbf{\textit{I'm interested in culture and architecture; suggest cities in Europe for a trip in autumn.}}} \\ 

\textbf{Baseline}: \\
\textcolor{BlueViolet}{1. I recommend Sibiu, Romania. I recommend Sibiu, Romania because it is a city rich in history and culture. The city has a well-preserved medieval old town with narrow streets and colorful buildings, which is perfect for exploring. The city also has a strong cultural identity, with many festivals and events throughout the year, including the famous Sibiu International Theatre Festival.   \\ ...} \\

\textbf{SAR}: \\
\textcolor{ForestGreen}{ I recommend Strasbourg, Sibiu, and Plovdiv. I recommend these cities because they have lower s-fairness scores, which indicates that they are more sustainable options for travel in November. Strasbourg is a historic city with a well-preserved city centre that is a UNESCO World Heritage Site. Sibiu is a charming city with a rich cultural heritage, featuring a mix of Gothic and Romanesque architecture. Plovdiv is a vibrant city with a rich history, offering a glimpse into Bulgaria's past. All three cities offer a unique cultural experience, and their lower s-fairness scores make them more sustainable options for travel in November.}
}

To effectively evaluate the pipeline, a diverse set of test queries reflecting various user preferences is essential, and recent research suggests using LLMs to create synthetic test data for different domains~\cite{li2023synthetic, bao2023synthetic}, including recommender systems~\cite{rahmani2024synthetic}.
We use state-of-the-art LLMs, including GPT-4 \cite{achiam2023gpt} and Gemini-1.5-Pro \cite{reid2024gemini}, to generate 200 diverse test queries about European travel destinations for different months and seasons, such as recommending winter destinations for hiking and skiing. 
Each of the 200 queries is passed to both models twice ---once using the basic system prompt (baseline) and once with the sustainability prompt and S-Fairness indicators (SAR), as illustrated in~\autoref{fig: prompt}. This results in a total of 800 responses for analysis.

As evident from~\autoref{fig: sample_prompt_response}, we can see a sample generated test prompt using GPT-4 and the corresponding responses for both baseline and SAR using \llama.
The top recommendation for SAR, Strasbourg, has the lowest S-Fairness indicator among the ten retrieved cities, indicating that the model effectively balances user preferences with sustainability concerns.

 \subsection{Evaluation Metrics} \label{section: metrics}

Since RAG systems often consist of multiple independent components, evaluating the system can be challenging, as different phases need to be evaluated.
For the scope of this paper, we only consider the final LLM-generated response for both baseline and SAR for evaluation. 
Our metrics are inspired by the RAGAS framework proposed by~\citet{es2023ragas}.
The research questions we aim to address through our evaluation, along with their updated results after further optimization (as presented in~\autoref{tab: results}), are discussed in the following subsections.

\begin{table}[]
    \centering
    \caption{
      Table comparing the performance of the \llama and \mistral models across 200 prompts, evaluating both baseline and SAR scenarios using different metrics.}
    \medskip
    \label{tab: results}
    \resizebox{\textwidth}{!}{%
    \begin{tabular}{|l|l|lllll|}
        \hline
        \multicolumn{1}{|c|}{\multirow{3}{*}{\textbf{Models}}} & \multicolumn{1}{c|}{\multirow{3}{*}{\textbf{Method}}} & \multicolumn{5}{c|}{\textbf{Evaluation Metrics}}                                                                                   \\ \cline{3-7} 
                                         & \multicolumn{1}{c|}{}                                 & \multicolumn{2}{c|}{\textbf{Relevance}}                          & \multicolumn{2}{c|}{\textbf{Sustainability (\%)}}             & \textbf{Faithfulness (\%)} \\ \cline{3-6}
                                         &                                                       & \multicolumn{1}{l|}{\gpt}  & \multicolumn{1}{l|}{\claude} & \multicolumn{1}{l|}{Accuracy}  & \multicolumn{1}{l|}{Frequency}   &                   \\ \hline
        \multirow{2}{*}{\llama}           & Baseline                                              & \multicolumn{1}{c|}{8.16\sd{1.78}} & \multicolumn{1}{c|}{6.29\sd{2.27}}   & \multicolumn{1}{c|}{-} & \multicolumn{1}{c|}{-}  & \multicolumn{1}{c|}{0}                 \\  
                                         & SAR                                                   & \multicolumn{1}{c|}{7.69\sd{1.73}} & \multicolumn{1}{c|}{5.50\sd{2.29}}  & \multicolumn{1}{c|}{10.5}   & \multicolumn{1}{c|}{42.5} & \multicolumn{1}{c|}{0}              \\ 
        \multirow{2}{*}{\mistral}         & Baseline                                              & \multicolumn{1}{c|}{3.85\sd{2.72}} & \multicolumn{1}{c|}{3.05\sd{2.45}}   & \multicolumn{1}{c|}{-} & \multicolumn{1}{c|}{-}  & \multicolumn{1}{c|}{14.0} \\ 
                                         & SAR                                                   & \multicolumn{1}{c|}{3.96\sd{2.65}} & \multicolumn{1}{c|}{2.97\sd{2.35}}   & \multicolumn{1}{c|}{7.5}  & \multicolumn{1}{c|}{36.5} & \multicolumn{1}{c|}{9.5} \\ \hline
    \end{tabular}%
    }
\end{table}

\subsubsection{\textbf{Answer Relevance}}  \label{section: relevance}

Answer Relevance measures how well the LLM-generated response answers the question~\cite{es2023ragas}. In situations where manual evaluation is infeasible, LLMs have previously been shown to perform remarkably well in judging answers in a human way when provided with a suitable grading scheme~\cite {zheng2024judging, wang2023chatgpt}. 

We employ two LLMs to judge Answer Relevance --- \gpt~\cite{achiam2023gpt} and \claude~\cite{anthropic2024claude} by instructing them to grade the answer on a scale of 0 - 10, where 0 means that the answer is completely irrelevant and 10 implies that the answer is relevant and completely answers the question\footnote{https://huggingface.co/learn/cookbook/en/llm\_judge}. We compute the mean and standard deviation (indicated by the \sd{values}) for both \llama and \mistral for the baseline and SAR, as shown in \autoref{tab: results}.

Responses generated by \llama have higher average scores than those generated by \mistral, regardless of the judge, indicating \mistral often struggles to provide answers of high quality. 
Furthermore, our observations reveal that the average scores exhibit consistency upon incorporating SAR, maintaining similar levels as before its inclusion. Notably, when evaluating \mistral alongside~\gpt, for SAR, there is even a marginal improvement in the average scores. This suggests that the introduction of SAR preserves the overall quality and relevance of the answers to the question while taking into consideration sustainability during response generation.

\subsubsection{\textbf{Sustainability}} \label{section: sustainability}

As discussed in \autoref{section: prompt_augmentation}, the main goal of SAR is to help models consider the societal impact of recommended cities during the reranking process. 
For our evaluation, we focus on the S-Fairness ranks of the retrieved cities, where a better rank indicates a lower S-Fairness indicator value. 
To extract the list of recommended cities, we tokenize the generated response and then compare it with the list of retrieved cities from the context. 
We measure the \textit{accuracy} of our methodology by measuring how often each model selects the city with the lowest sustainability score as its top choice.
Our results, as described under "Sustainability" in~\autoref{tab: results}, show that \llama recommends the most sustainable city as a top candidate 10.5\% of the time, compared to \mistral at 7.5\%.

Since SAR instructs the LLMs to rerank the retrieved context and select the top 3 cities, we also measure the \textit{frequency} with which the model's top choice is the most sustainable among the recommended cities. This approach helps us account for scenarios where cities more aligned with user preferences may not be the most sustainable but ensures sustainability is still a factor in ranking the top choices.
Here, the results are more favorable for both \llama and \mistral, with the top choice having the lowest relative S-Fairness rank 42.5\% and 36.5\% of the time, respectively. This suggests that while the models may not always prioritize the most sustainable cities, sustainability still plays a significant role in their reranking process.

\subsubsection{\textbf{Model Agreement \& Faithfulness}} \label{section: agreement}

We also investigate whether the models agree in their responses when given the same prompts and context. Our analysis shows that the overall agreement between the baseline models is low, with both models recommending the same set of cities only 4\% of the time. However, the partial agreement is significantly higher, at 49\%, indicating that the models recommend at least one common city almost half the time.

Furthermore, we observe an increase in partial agreement (60.5\%) when SAR is included in the prompt.  The S-Fairness indicator introduced by SAR encourages the models to consider cities with the lowest scores, which may lead to more frequent alignment in their recommendations compared to the baseline method. However, the total agreement decreases upon including SAR (1\%).

Evaluating the faithfulness of LLM-generated text has largely relied on the availability of reference answers that can be compared with the candidates to compute similarity or conditional probability-based metrics~\citep{zhang_bertscore_2020, yuan2021bartscore}. However, in the context of RAG, faithfulness can be defined in terms of the generated answer and the retrieved context~\cite{es2023ragas}. We compute the faithfulness by counting the prompts with out-of-context (OC) responses, which indicate complete model hallucination. 

Our results, listed under "Faithfulness" in~\autoref{tab: results} show that \llama performs exceptionally well, with no OC responses in the baseline model, outperforming \mistral, which hallucinates 14\%  of the time. Introducing SAR reduces hallucinations in \mistral to 9.5\%, further supporting its effectiveness. With \llama, neither the baseline nor the SAR-enabled models recommend any OC responses.

\section{Conclusion and Future Work} \label{section: conclusion}

This paper presents a novel approach to enhancing TRS by incorporating sustainability metrics during the prompt augmentation phase of an RAG pipeline, ensuring that sustainability is prioritized in the recommendation process. 
Our findings using two popular open-source models --- \llama and \mistral indicate that the addition of the Sustainability Augmented Reranking (SAR) generally matches or enhances model performance, without compromising answer quality.

Future research could further refine this approach by expanding the knowledge base to include more diverse, real-time datasets and exploring additional sustainability metrics, such as carbon footprint and local economic impact, to provide a more holistic assessment of sustainable travel recommendations.
Examining the effects of various prompts and incorporating popularity and seasonality indices into the context, or using S-Fairness ranks rather than absolute values, could provide valuable insights into LLM performance. 
Ranking sustainability during the retrieval phase may also enhance alignment with sustainability goals, ensuring that potential sustainable destinations are not overlooked.

Currently, there is no user information available, posing a cold-start problem. Future work could address this by developing a conversational recommender system that tailors recommendations based on user preferences and profiles. Moreover, response generation currently occurs in a zero-shot setting; future iterations might explore few-shot learning and in-context learning to improve recommendation relevance and alignment with sustainability goals.
These refinements could potentially improve the effectiveness of LLMs in delivering recommendations that are both relevant and aligned with sustainability objectives.

{
\makeatletter
\renewcommand{\bibsection}{%
   \section*{\refname}%
   \@mkboth{\MakeUppercase{\refname}}{\MakeUppercase{\refname}}%
}
\makeatother
\bibliographystyle{apalike}
\bibliography{references}
}
\appendix

\end{document}